\documentclass[prl,preprint,showpacs,superscriptaddress]{revtex4}
\usepackage{latexsym}
\usepackage{amsfonts}
\usepackage{graphicx}
\usepackage{epsfig}
\usepackage{color}

\begin{document}

\title{Glass transition in biomolecules and the liquid-liquid critical
point of water}

\date{LS10901 last revised 9:30  pm ; printed: \today }

\author{P. Kumar}
\affiliation{Center for Polymer Studies and Department of Physics, Boston
 University,~Boston, MA 02215 USA}
\author{Z. Yan }
\affiliation{Center for Polymer Studies and Department of Physics, Boston
 University,~Boston, MA 02215 USA}
\author{L. Xu}
\affiliation{Center for Polymer Studies and Department of Physics, Boston
 University,~Boston, MA 02215 USA}
\author{M. G. Mazza}
\affiliation{Center for Polymer Studies and Department of Physics, Boston
 University,~Boston, MA 02215 USA}
\author{S. V. Buldyrev}
\affiliation{Department of Physics,~Yeshiva
 University, 500 West 185th Street,~New York, NY 10033 USA}
\affiliation{Center for Polymer Studies and Department of Physics, Boston
 University,~Boston, MA 02215 USA}
\author{S.-H. Chen}
\affiliation{Nuclear
 Science and Engineering Department,~Massachusetts Institute of Technology,
 Cambridge, MA 02139 USA}
\author{S. Sastry}
\affiliation{Javaharlal Nehru Center for Advanced
 Scientific Research, Jakur Campus, Bangalore, 560061, India}
\author{H.~E.~Stanley}
\affiliation{Center for Polymer Studies and Department of Physics, Boston
 University,~Boston, MA 02215 USA}

\bigskip
\bigskip

\begin{abstract}
Using molecular dynamics simulations, we investigate the relation
between the dynamic transitions of biomolecules (lysozyme and DNA) and
the dynamic and thermodynamic properties of hydration water. We find
that the dynamic transition of the macromolecules, sometimes called a
``protein glass transition'', occurs at the temperature of dynamic
crossover in the diffusivity of hydration water, and also coincides with
the maxima of the isobaric specific heat $C_P$ and the temperature
derivative of the orientational order parameter.   We relate these
findings to the hypothesis of a liquid-liquid critical point in
water. Our simulations are consistent with the possibility that the
protein glass transition results from crossing the Widom line, which is
defined as the locus of correlation length maxima emanating from the
hypothesized second critical point of water.

\end{abstract}

\maketitle

Both experiments and computer simulation studies have shown that
hydrated proteins undergo a ``glass-like'' transition near $200$~K
\cite{Ringe,Hart,Lee,Rasmussen_Nature_1992,Doster_nature_1989}, above
which proteins exhibit diffusive motion, and below which the proteins
are trapped in harmonic modes. An important issue is to determine the
effects of hydration water on this dynamical
transition~\cite{Vitkup_nature_struct_2000,Tarektobias,Bellisent,sokolov99,norberg96}.
Experiments and computer simulations suggested that when a protein is
solvated, the protein glass transition is strongly coupled to the
solvent, leading to the question of whether the protein glass transition
is directly related to a dynamic transition in the surrounding
solvent~\cite{Smith_BiophysJ_2003}.

It has been hypothesized that the deeply supercooled region of the
phase diagram of liquid water may contain a first order phase
transition between a high density liquid (HDL) and a low density
liquid (LDL)~\cite{poole_nature}. This line of phase transition has a
negative slope in the P-T phase diagram and terminates with a critical
point, $C_2$~\cite{poole_nature,brov05,brov05-2,PNAS}, which is
located at $T_{C2}=200 \pm 20$~K and $P_{C2}=250 \pm
100$~MPa~\cite{note1}. Upon crossing the first order phase transition
line above the critical pressure, the thermodynamic state functions
change discontinuously. Below the critical pressure they rapidly but
continuously change upon cooling.

Computer simulations of the TIP5P and ST2 models show that
response functions such as isobaric specific heat $C_P$ and isothermal
compressibility have maxima as functions of temperature if the system
is cooled isobarically at $P<P_{C2}$~\cite{PNAS}. The loci of these maxima
asymptotically approach one another~\cite{poole2} as the critical
point is approached, because response functions become expressible
in terms of the correlation length which diverges at the critical
point. The locus of the correlation length maxima is called the Widom
line.

Experimental studies of supercritical water~\cite{Anisimovbook} indeed
show that various response functions have sharp maxima in the region
of the phase diagram above the liquid-vapor critical point $C_1$, but
no direct experimental indication of a liquid-liquid critical point
$C_2$ had been available due to unavoidable crystallization of bulk
water. It was found that water remains unfrozen in hydrophilic
nanopores for $T>200$~K~\cite{mori99,Chen_PRL_2005}. Moreover when
cooled at constant pressure for $P<P_{C2}$ the dynamics changes from
non-Arrhenius to Arrhenius at $T=T_\times(P)$. The line $T_\times(P)$
is located in the range of temperatures between $200-230$~K and has a
negative slope in the P-T phase diagram. Computer simulations suggest
that this line may be associated with $T_W(P)$, the Widom line, near
which the local dynamic characteristics must rapidly change from those
resembling the properties of HDL at high temperature to those of LDL
at low temperature~\cite{PNAS}.

Here we explore the hypothesis~\cite{Chen_Nature_2006} that the
observed glass transition in biomolecules is related to the
liquid-liquid phase transition using molecular dynamics (MD)
simulations.  Specifically, we study the dynamic and thermodynamic
behavior of lysozyme and DNA in hydration TIP5P
water, by means of the software
package GROMACS~\cite{lindahl_J_mol_mod_2001} for (i) an orthorhombic
form of hen egg-white lysozyme~\cite{ortholyme} and (ii) a Dickerson
dodecamer DNA~\cite{drew86} at constant pressure $P=1$~atm, several
constant temperatures $T$, and constant number of water molecules $N$
(NPT ensemble) in a simulation box with periodic boundary
conditions. We first allow the system to equilibrate at constant
temperature and pressure using the Berendsen method. This initial
equilibration is followed by a long production run during which we
calculate the dynamic and static properties. Equilibration times vary
for different temperatures from a few ns for high temperatures to as
much as $40$~ns for low temperatures. The MD for DNA was performed
using the Amber force field~\cite{sorin05}.  For lysozyme simulations,
the system consists of a single protein in the native conformation
solvated in $N=1242$ TIP5P water
molecules~\cite{tip5p,Yamada_PRL_2002,brov05-2}. These simulation
conditions correspond to a ratio of water mass to protein mass of
$1.56$. The DNA system consists of a single DNA helix with 24
nucleotides solvated in $N=1488$ TIP5P water molecules, which
corresponds to an experimental hydration level of $3.68$.

\begin{figure}
\centerline{
\includegraphics[width=8cm]{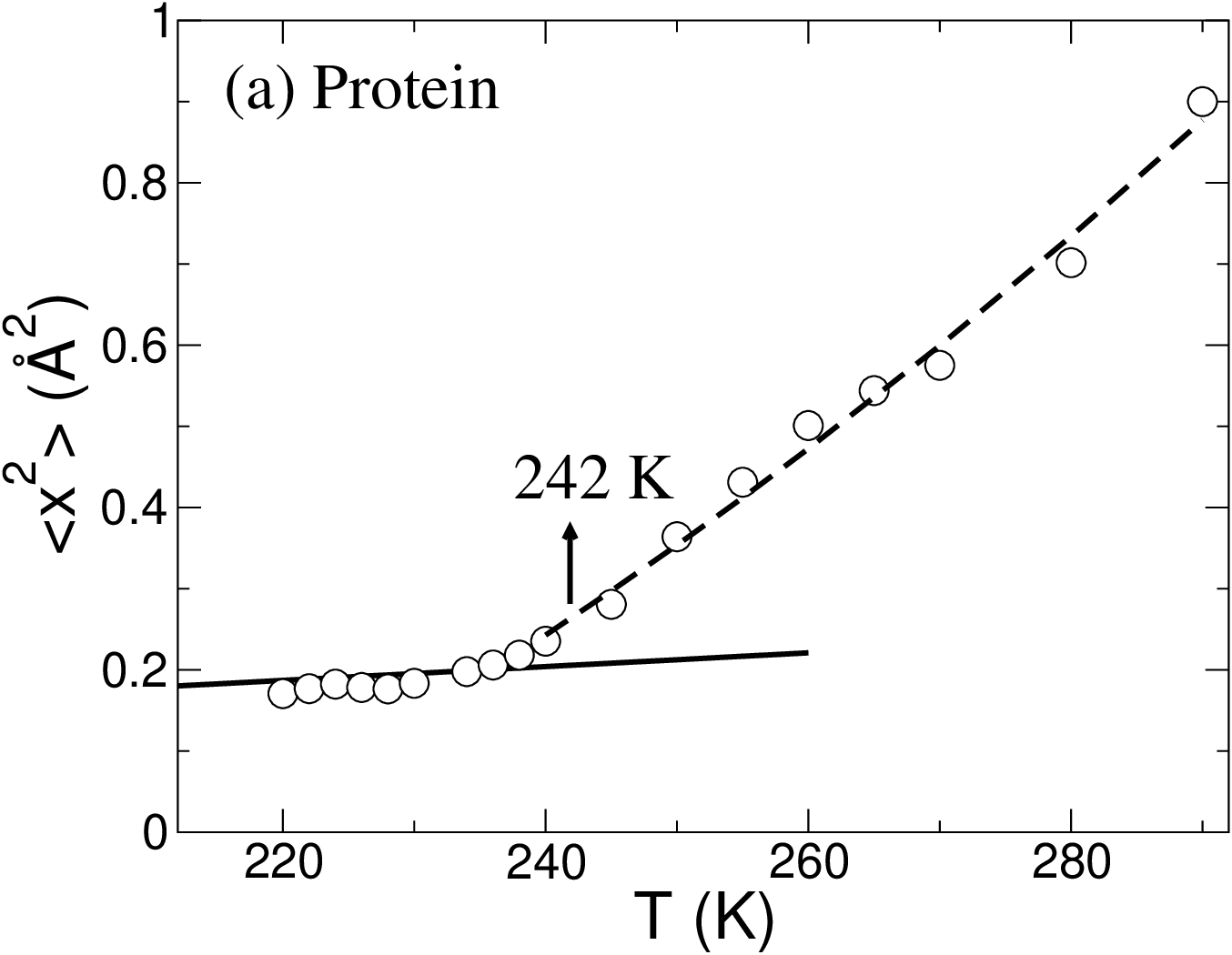}
\includegraphics[width=8cm]{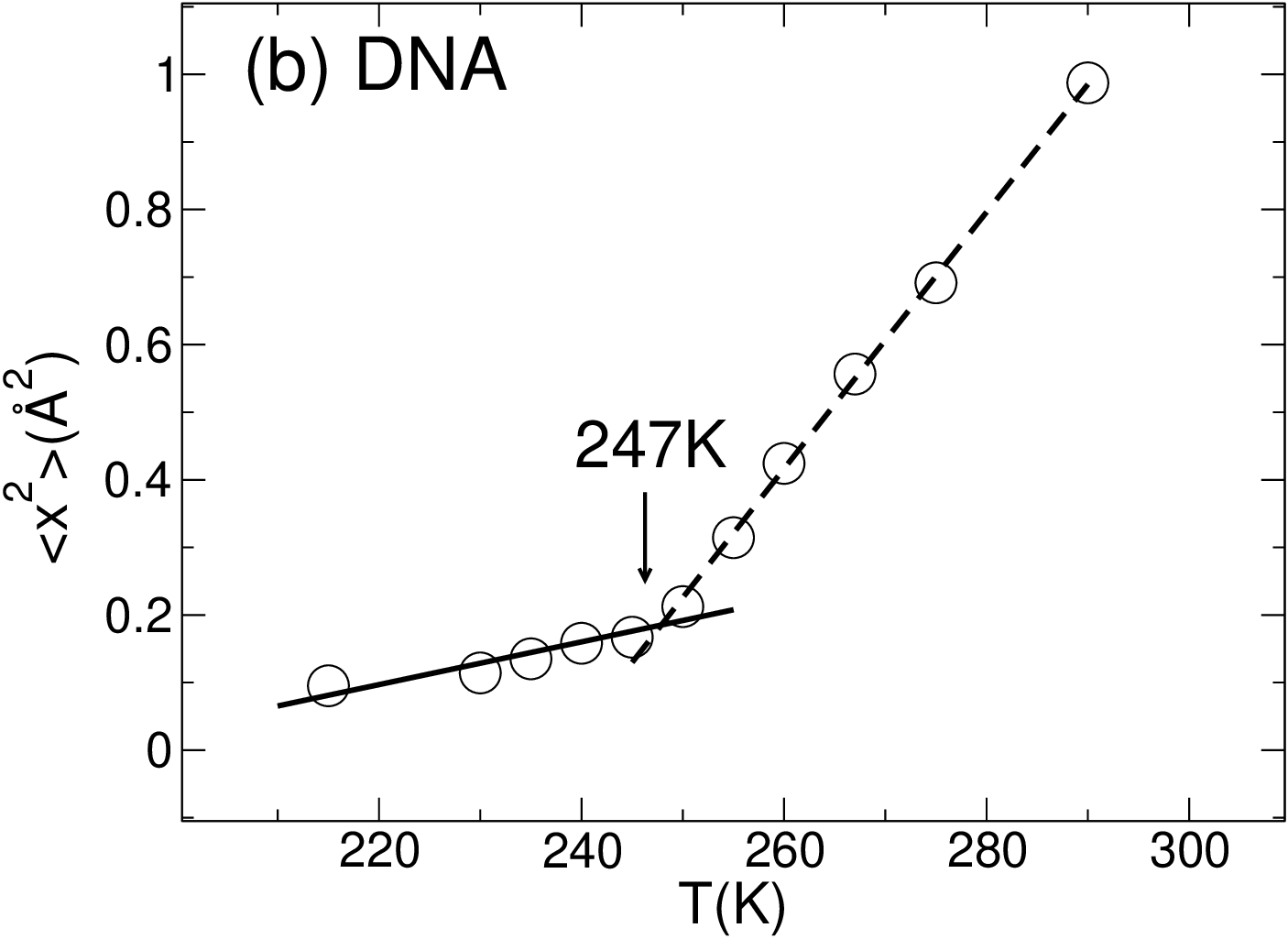}
}
\centerline{
\includegraphics[width=8cm]{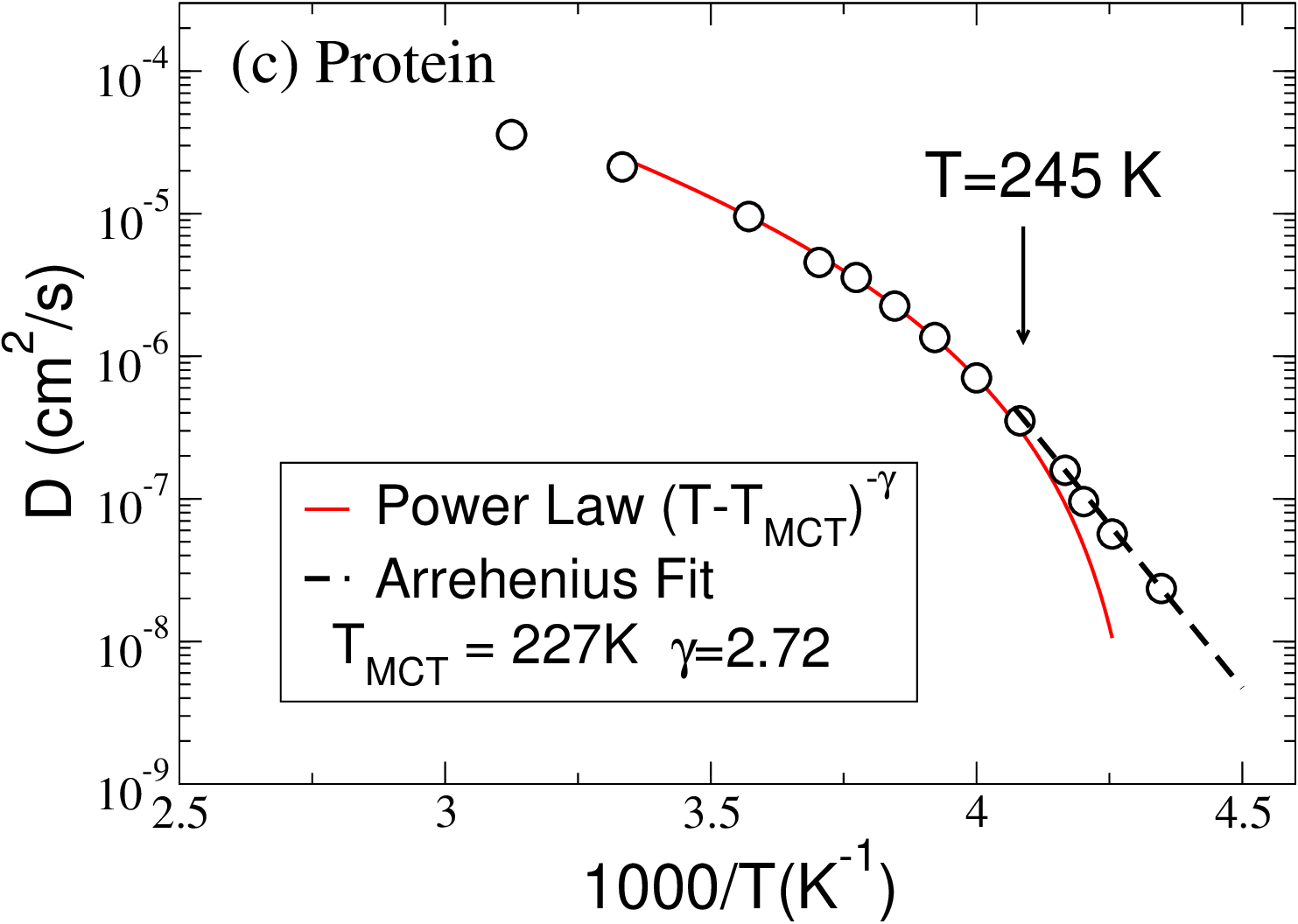}
\includegraphics[width=8cm]{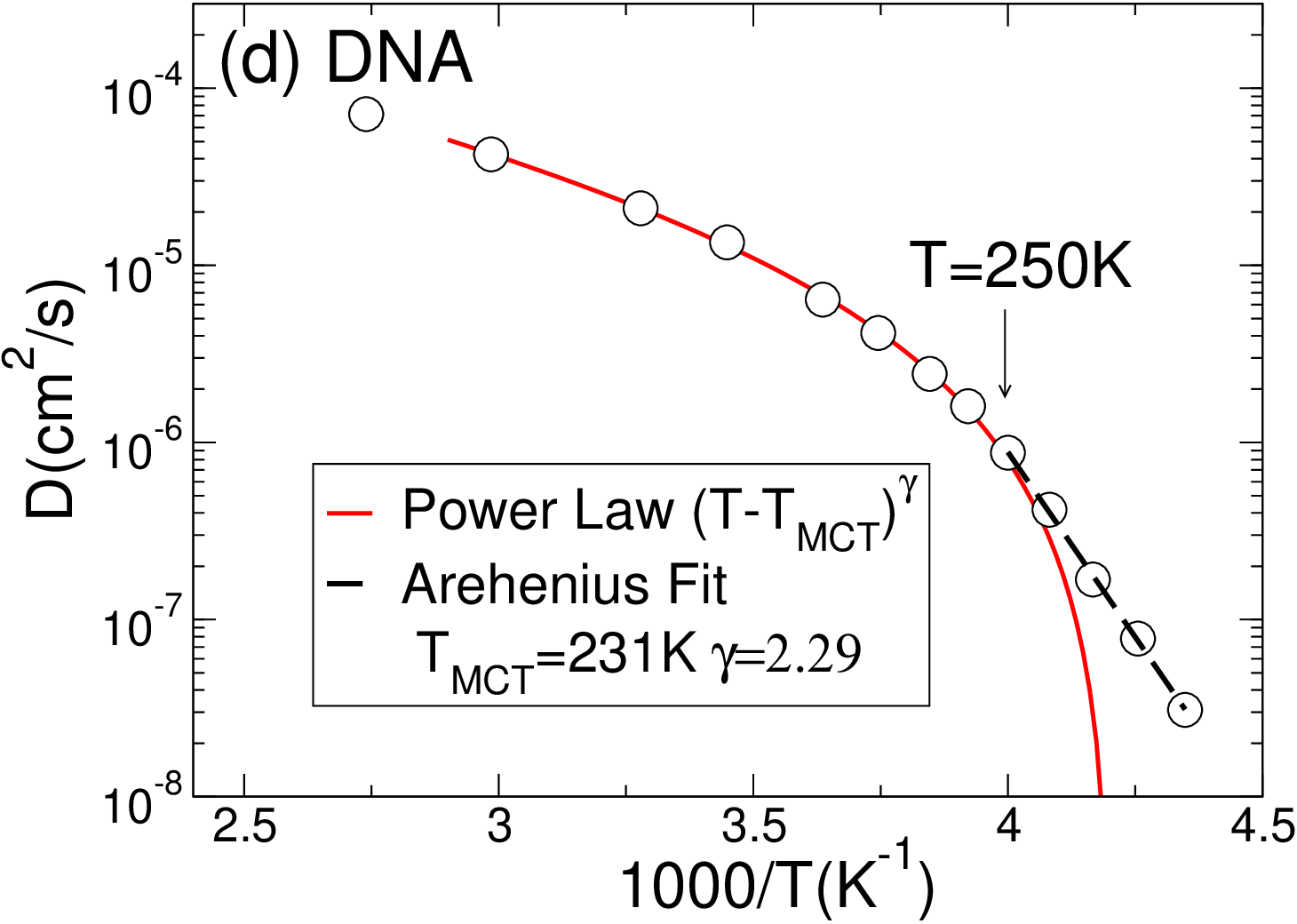}
}
\caption{Mean square fluctuation  of (a) lysozyme, and (b) DNA
 showing that there is a transition around $T_{\rm p}\approx242\pm10$~K
 for lysozyme and around $T_{\rm p}\approx247\pm10$~K for DNA. For very
 low $T$ one would expect a linear increase of $<x^2>$ with $T$, as a
 consequence of harmonic approximation for the motion of residues. At
 high T, the motion becomes non-harmonic and we fit the data by a
 polynomial. We determine the dynamic crossover temperature $T_{\rm p}$ from
 the crossing of the linear fit for low T and the polynomial fit for high
 T. We determine the error bars by changing the number of data points
 in the two fitting ranges.
Diffusion constant of hydration water surrounding (c)
 lysozyme, and (d) DNA shows a dynamic transition from a power law
 behavior to an Arrhenius behavior at $T_{\times}\approx 245\pm10$~K for
 lysozyme and $T_{\times}\approx 250\pm10$~K for DNA, around the same temperatures
 where the behavior of $<x^2>$ has a crossover.}

\label{fig:fluctuat}
\end{figure}

The simulation results for the mean square fluctuations $<x^2>$ of
protein are shown in Fig.~\ref{fig:fluctuat}(a). We calculate the mean
square fluctuations $<x^2>$ of protein from the equilibrated
configurations, first for each atom over $1$~ns, and then averaged
over the total number of atoms in the protein. We find that
$<x^2>$~[Fig.~\ref{fig:fluctuat}(a)] changes its functional form below
$T_{\rm p}\approx 242\pm10$~K.
 Moreover, upon cooling, the diffusivity of hydration water exhibits a
dynamic crossover from non-Arrhenius~\cite{note1a} to Arrhenius behavior
at the crossover temperature $T_\times\approx 245\pm10$~K
[Fig.~\ref{fig:fluctuat}(c)]. 
 A similar temperature dependence of diffusivity of bulk TIP5P water
was observed in Ref.~\cite{PNAS}. The coincidence of $T_\times$ with
$T_{\rm p}$ within the error bars indicates that the behavior of the
protein is strongly coupled with the behavior of the surrounding
solvent, in agreement with recent experiments
~\cite{Chen_Nature_2006}. Note that $T_\times$ is much higher than the
glass transition temperature estimated for TIP5P as $T_g=215$K
\cite{brov05-2}. Thus this crossover is not likely to be related to
the glass transition in water.  Here we will explore the possibility
that instead it is related to a change in the properties of protein
hydration water.

\begin{figure}
\centerline{
\includegraphics[width=8cm]{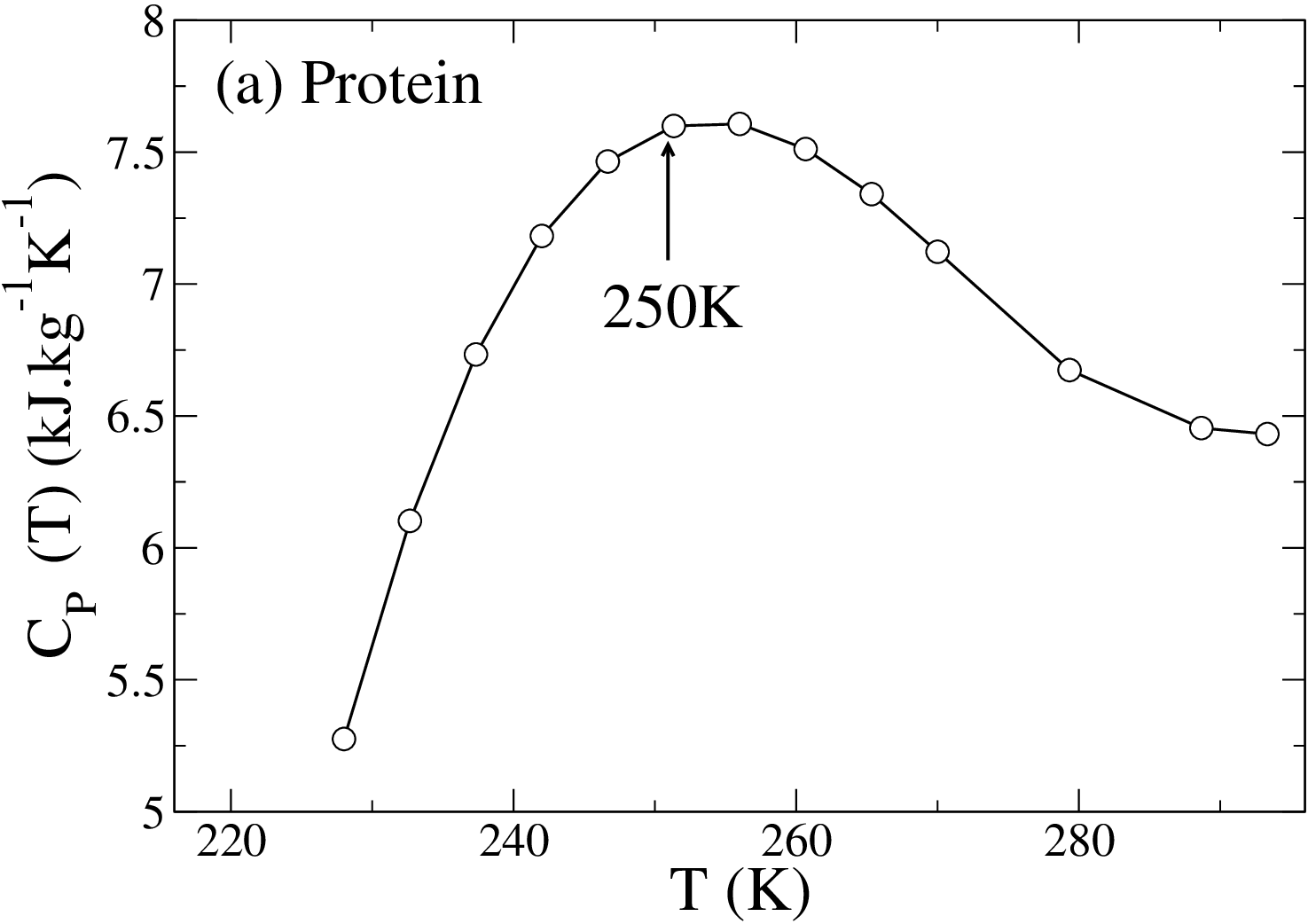}
\includegraphics[width=8cm]{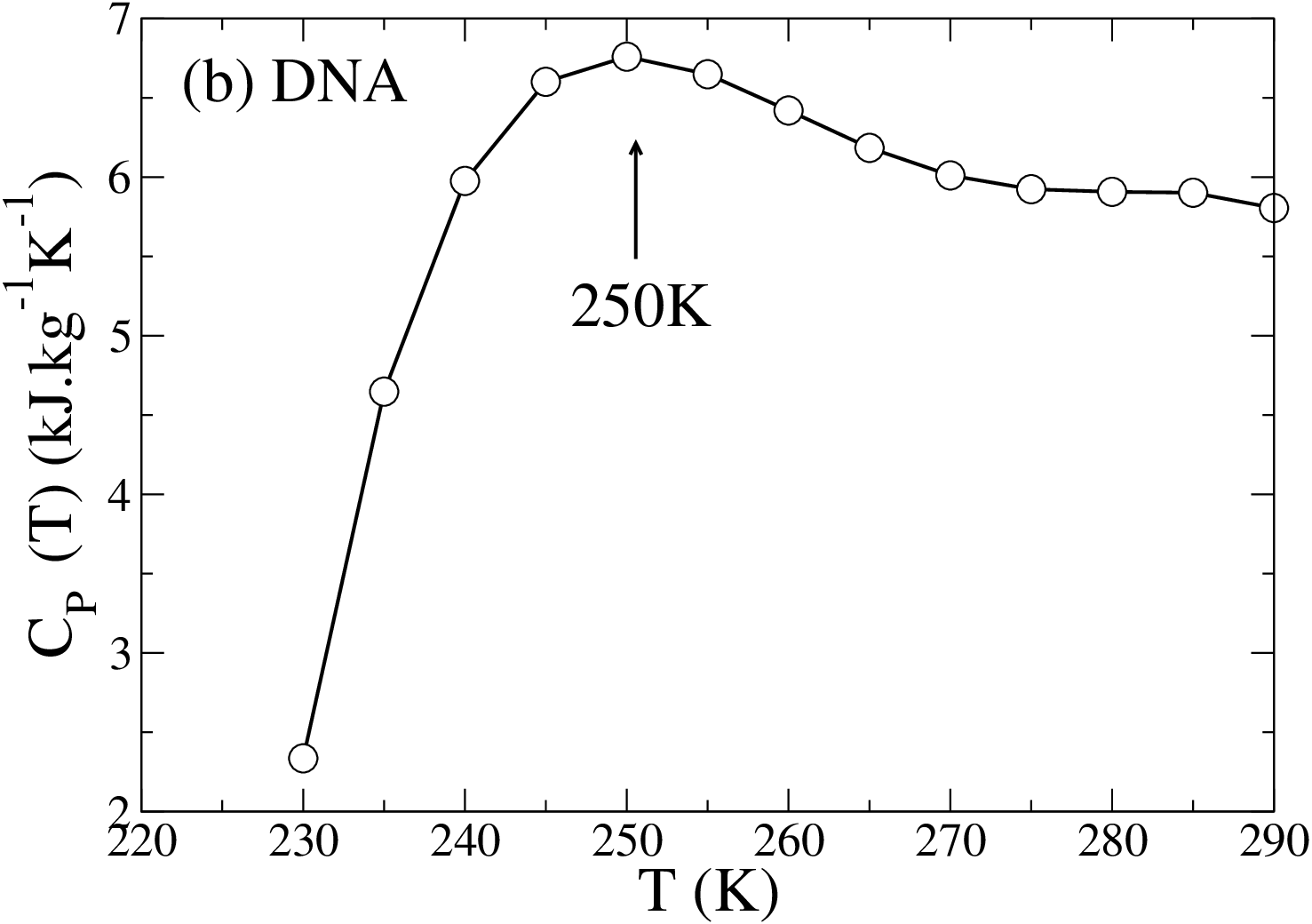}
}
\caption{The specific heat of the combined system (a) lysozyme and
 water, and (b) DNA and water, display maxima at $250\pm10$~K and
 $250\pm12$~K respectively, which are coincident within the error bars
 with the temperature $T_{\rm p}$ where the crossover in the behavior of
 $<x^2>$ is observed.}
\label{fig:dyn}
\end{figure}

We next calculate $C_P$ by numerical differentiation of the total
enthalpy of the system (protein and water) by fitting the simulation
data for enthalpy with a fifth order polynomial, and then taking the
derivative with respect to $T$. Figure~\ref{fig:dyn}(a) displays a
maximum of $C_P(T)$ at $T_{\rm W}\approx 250\pm10$~K for the case of
lysozyme-water system~\cite{note7}. The fact that $T_{\rm p} \approx
T_{\times} \approx T_{\rm W}$ is evidence of the correlation between
the changes in protein fluctuations [Fig.~\ref{fig:fluctuat}~(a)] and
the hydration water thermodynamics [Fig.~\ref{fig:dyn}~(a)]. Thus our
results are consistent with the possibility that the protein glass
transition is related to the Widom line (and hence to the hypothesized
liquid-liquid critical point).  Crossing the Widom line corresponds to
a continuous but rapid transition of the properties of water from
those resembling the properties of a local HDL structure for $T>T_{\rm
W}(P)$ to those resembling the properties of a local LDL structure for
$T<T_{\rm W}(P)$~\cite{PNAS, Chen_PRL_2005}. A consequence is the
expectation that the fluctuations of the protein residues in
predominantly LDL-like water (more ordered and more rigid) just below
the Widom line should be smaller than the fluctuations in
predominantly HDL-like water (less ordered and less rigid) just above
the Widom line.

\begin{figure}
\centerline{
\includegraphics[width=7cm]{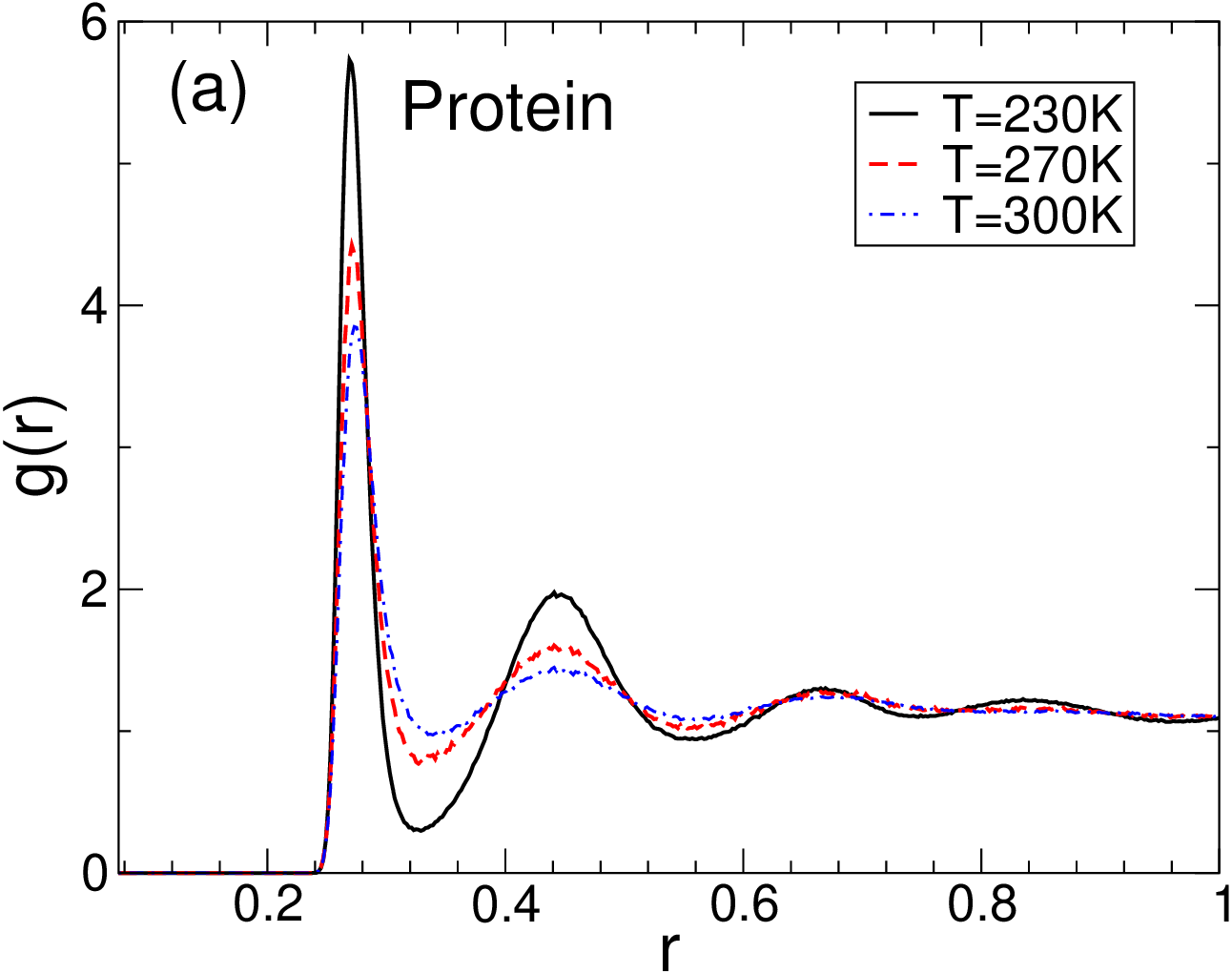}
\includegraphics[width=7cm]{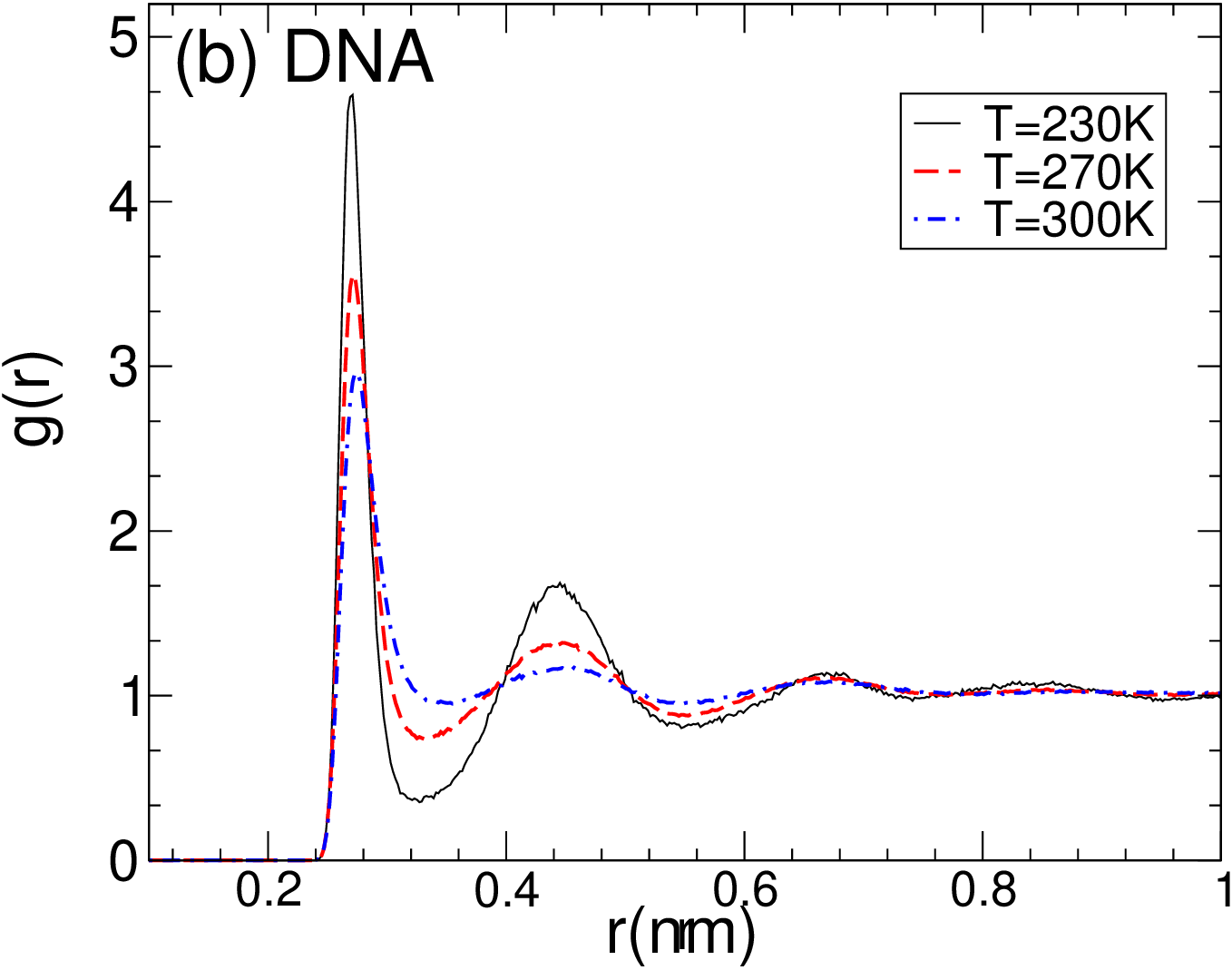}
}
\centerline{
\includegraphics[width=7cm]{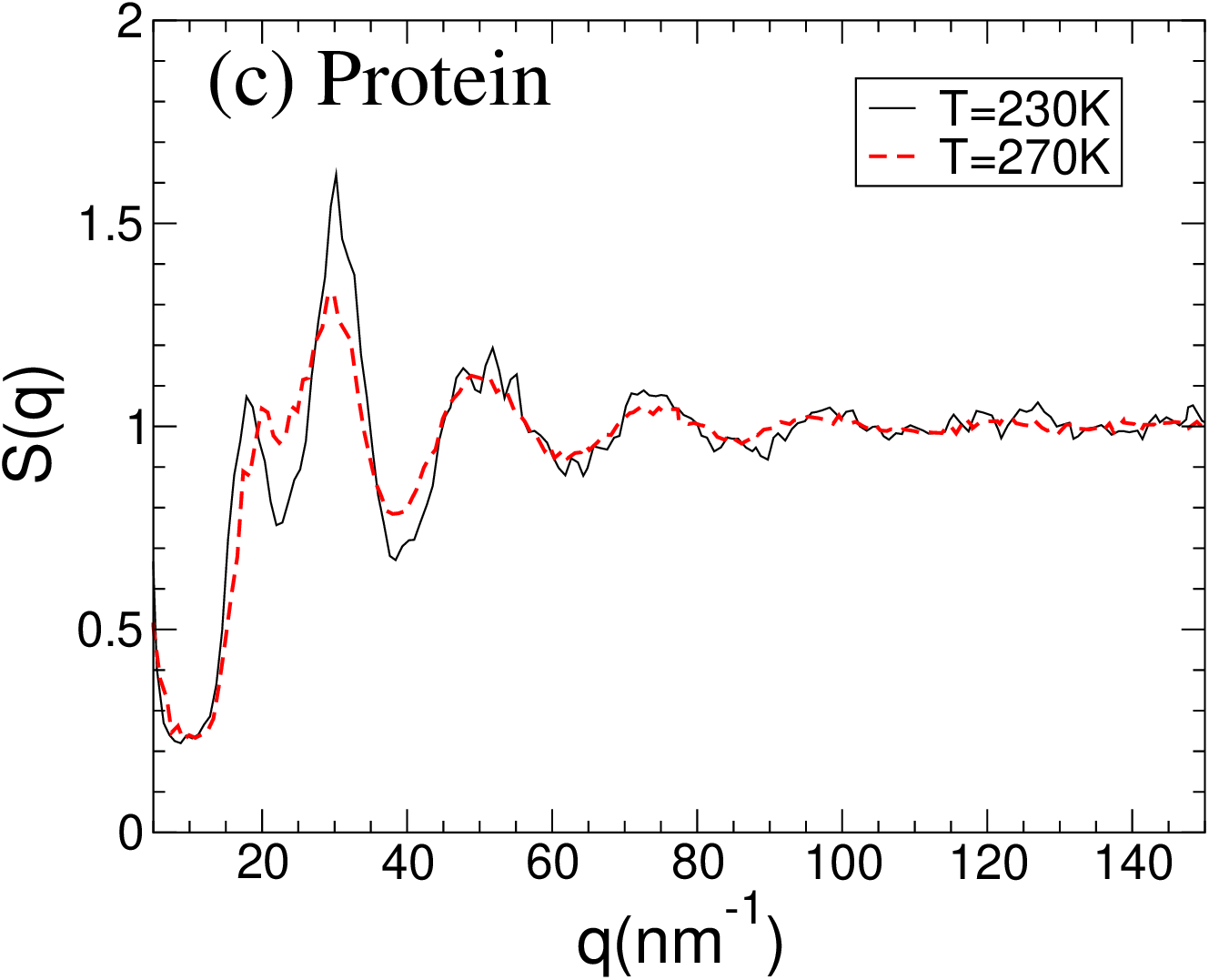}
\includegraphics[width=7cm]{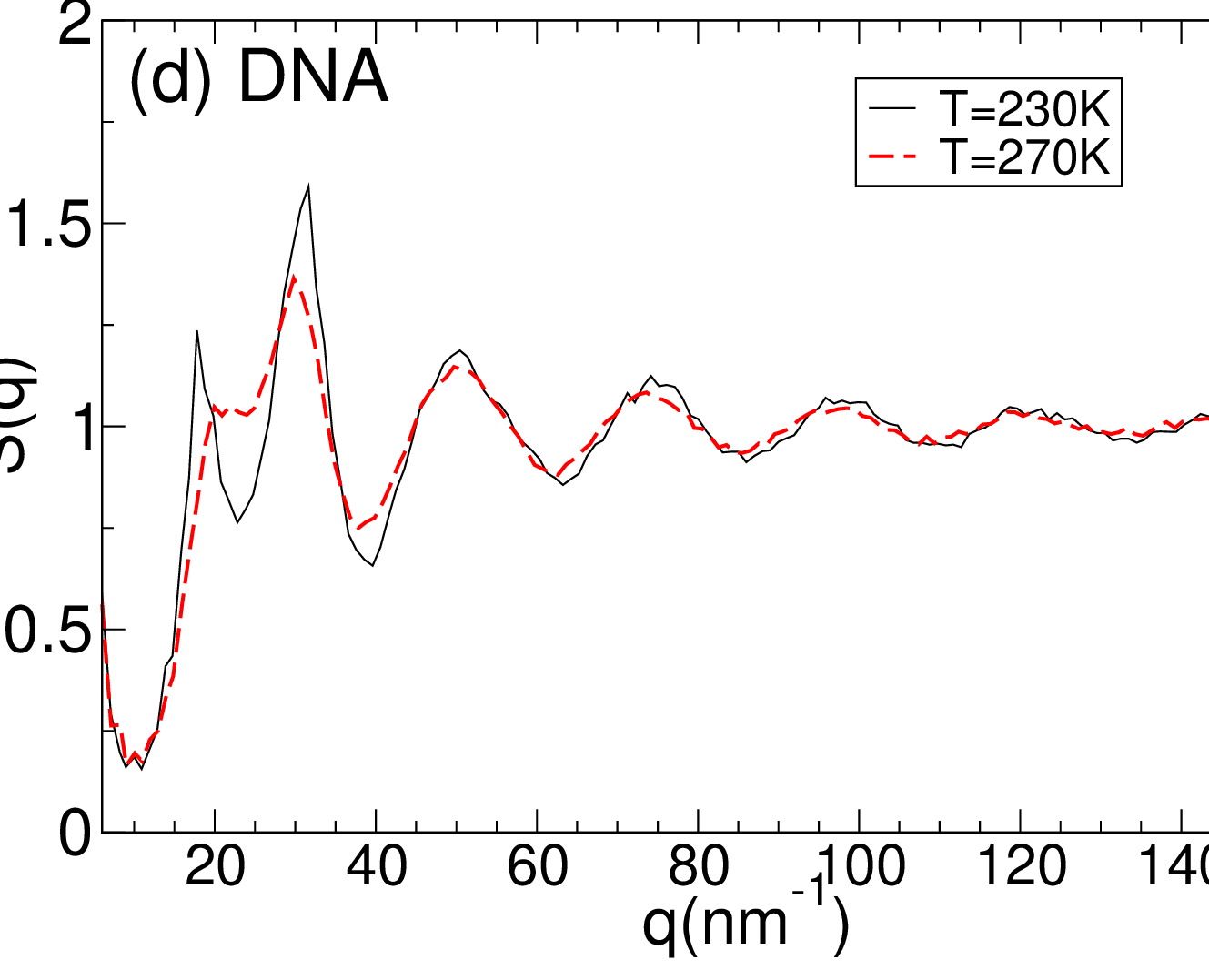}
}
\centerline{
\includegraphics[width=7cm]{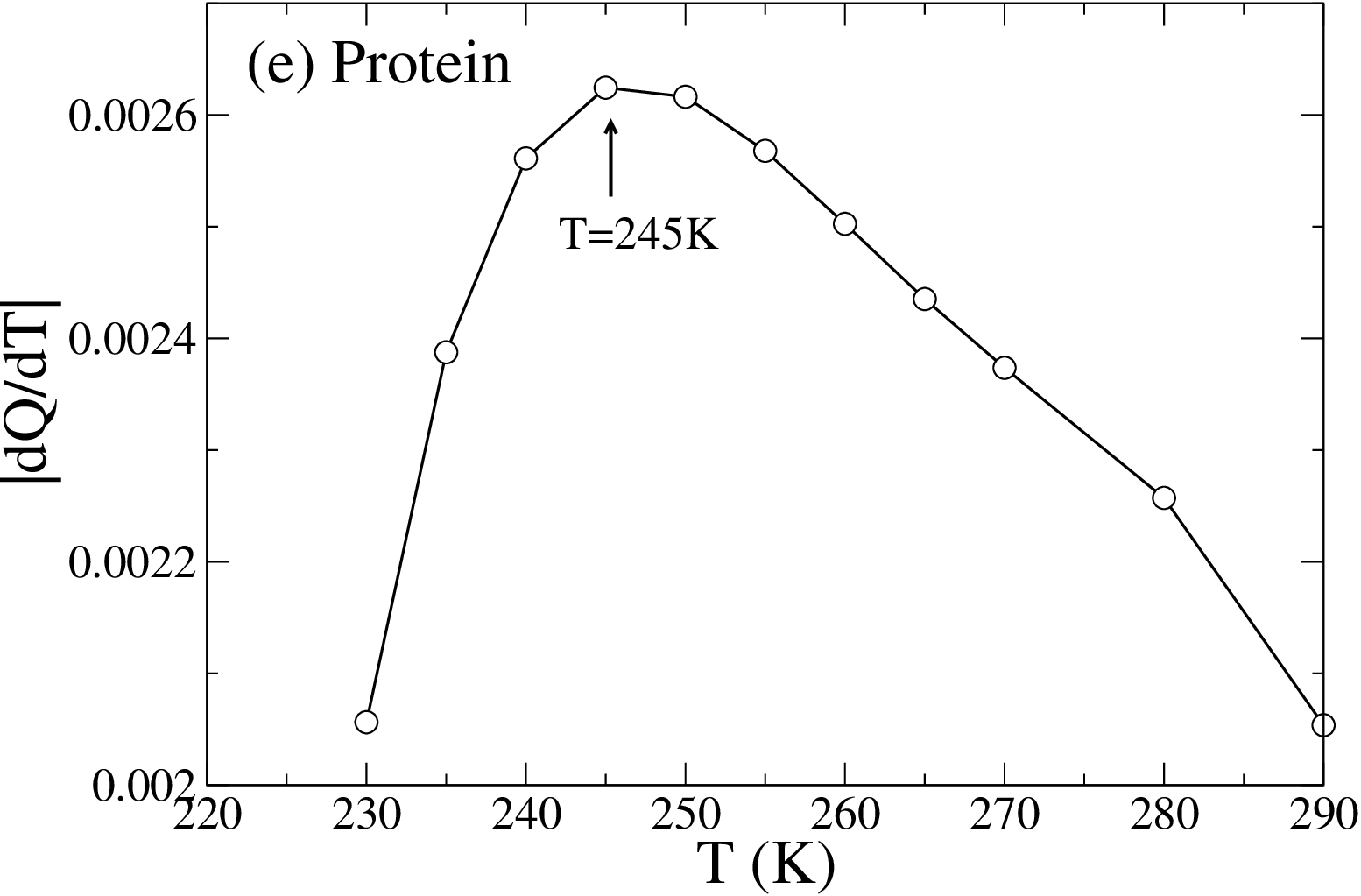}
\includegraphics[width=7cm]{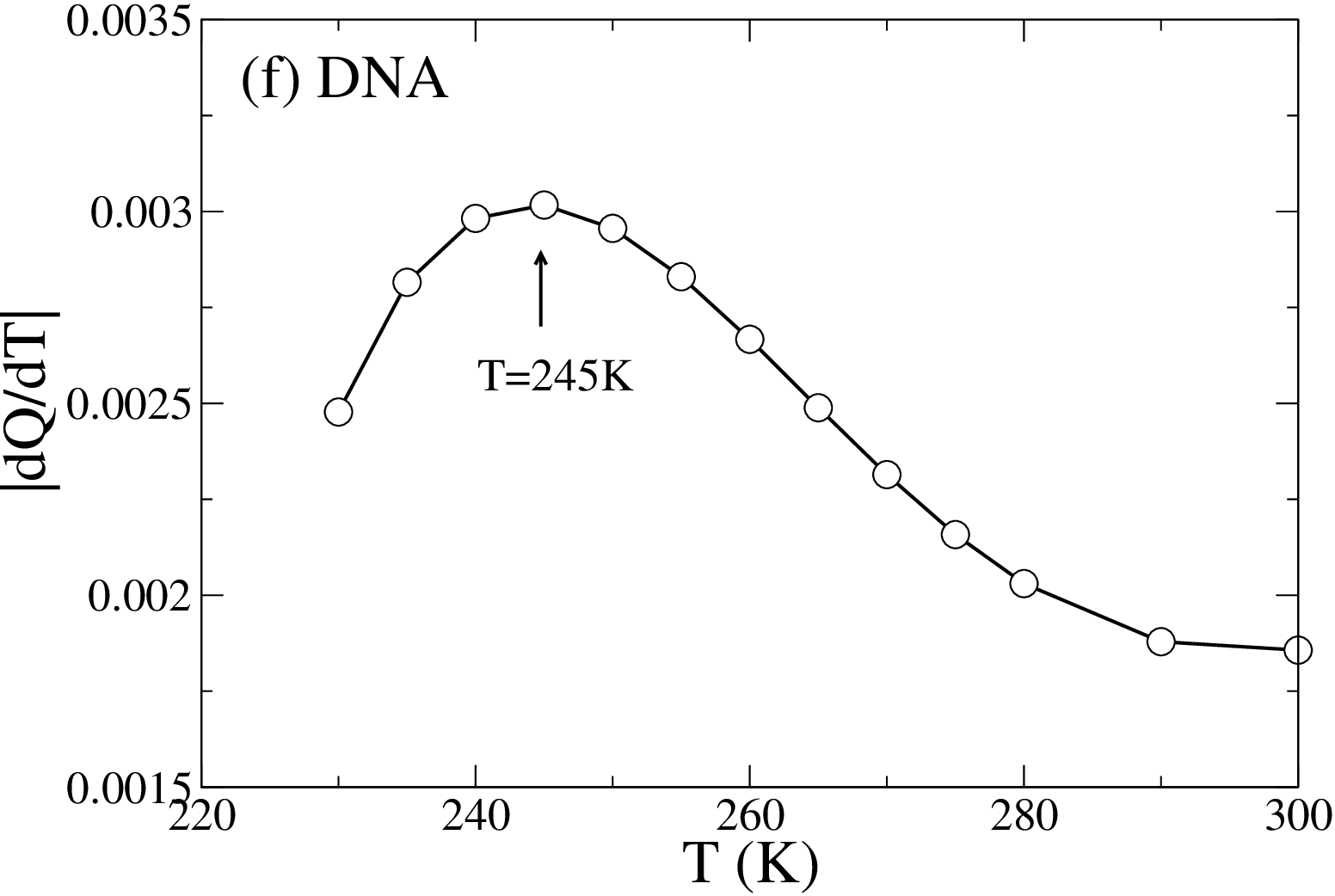}
}
\caption{ (Color online) Oxygen-Oxygen pair correlation function
 $g(r)$ for (a) lysozyme hydration water, and (b) DNA hydration water,
 on crossing the Widom line from the HDL side ($T=270$~K, $300$~K) to
 the LDL side ($T=230$~K). Structure factor of hydration water
 surrounding (c) lysozyme, and (d) DNA on two sides of the the Widom
 line. Upon crossing the Widom line, the local structure of water
 changes from more HDL-like to more LDL-like, reflected in the sharper
 and more prominent first peak. The first peak associated with the
 hydrogen bond distance also moves to small wave vectors, suggesting a
 change from the HDL to the LDL-like local structure of water at low
 temperatures. Derivative with respect to temperature of the local
 tetrahedral order parameter $Q$ for (e) lysozyme and (f) DNA
 hydration water. A maximum in $|dQ/dT|$ at Widom line temperature
 suggests that the rate of change of local tetrahedrality of hydration
 water has a maximum at the Widom line.}

\label{fig:structure}
\end{figure}

To test this interpretation, we analyze the structure of hydration water
on the two sides of the Widom line. Fig.~\ref{fig:structure}(a) shows
the oxygen-oxygen radial distribution function $g(r)$ on two sides of
the Widom line for lysozyme hydration water. The first peak of $g(r)$ on
the low temperature (T=$230$~K) side is sharper and the first minimum is
shallower compared to the $g(r)$ on the high temperature (T=$270$~K,
$300$~K) side of the Widom line, suggesting that water is more
structured on the low temperature side. Further, we calculate the
structure factor $S(q)$ of lysozyme hydration water
[Fig.~\ref{fig:structure}(c)]. The first peak of the structure factor
associated with the hydrogen bond is very sharp and pronounced, for
$T<T_{\rm W}(P)$, it is diminished and moves to larger wave vectors for
$T>T_{\rm W}(P)$, consistent with a LDL-like local structure for
$T<T_{\rm W}(P)$ and a HDL-like local structure for $T>T_{\rm
W}(P)$. 



Previous simulations~\cite{norberg96} and experiments~\cite{sokolov99}
suggest a ``glass-like'' transition of DNA around temperature
$230$~K. Hence to test if the dynamic crossover depends on the solute,
we performed a parallel study of the DNA Dickerson
dodecamer~\cite{drew86}. We find that fluctuations~\cite{note2} of the
DNA molecule [Fig.~\ref{fig:fluctuat}(b)] change their behavior
approximately at the same temperature as lysozyme, with
$T\approx247\pm10$~K.  The dynamic crossover in the hydration water upon
cooling from non-Arrhenius to Arrhenius behavior takes place at
$T_\times\approx250\pm10$~K [Fig.~\ref{fig:fluctuat}(d)].
 For DNA hydration water system $C_P$ has a maximum at
$T\approx250\pm12$~K, similar to the case of protein (see
Fig.~\ref{fig:dyn}(b))~\cite{note3}.  Fig.~\ref{fig:structure}(b)
and Fig.~\ref{fig:structure}(d) show $g(r)$ and $S(q)$ for the DNA
hydration water~\cite{note4}.  Further to describe the quantitative
changes in structure of hydration water, we calculate the local
tetrahedral order parameter $Q$~\cite{erringtonQ} for hydration water
surrounding lysozyme and DNA. Figs.~\ref{fig:structure}(e) and
~\ref{fig:structure}(f) show that the rate of increase of $Q$ has a
maximum at $245\pm10$~K for lysozyme and DNA hydration water
respectively; the same temperatures where we find a crossover in the
behavior of mean square fluctuations and a change in the behavior of
the dynamics of hydration water.

The quantitative agreement of the results for DNA and lysozyme suggests
that it is indeed the changes in the properties of hydration water
that are responsible for the changes in dynamics of the protein and
DNA biomolecules.  Our results are in qualitative agreement with
recent experiments on hydrated protein and DNA~\cite{Chen-PRL-06}
which found the crossover in side-chain fluctuations at $T_{\rm
p}\approx225$~K .

Note Added in Proof: After this work was submitted, we learned of
interesting parallel work on silicon, which also interprets structural
change in $g(r)$ and $S(q)$ as crossing the Widom line \cite{morishita}.

We thank C. A. Angell, L. Cruz, P. G. Debenedetti, L. Liu, P. J. Rossky,
and B. Urbanc for helpful discussions of both experimental and
computational work, and NSF for financial support.

\newpage

\newpage

\newpage

\begin{thebibliography}{100}
\bibitem{note1} The situation is rather complex. Many different
 scenarios have been proposed in the literature, including a
 scenario with multiple liquid-liquid critical points~\cite{brov05,brov05-2}.
\bibitem{note1a} To fit the data, we use a prediction of the ``mode
coupling theory'' (MCT)~\cite{gotze99} at high $T$ with parameters
$T_{MCT}\approx227$ and $\gamma\approx2.72$
\bibitem{note7}
The error bars in the enthalpy calculation do
not exceed 1\%. To assess the error bar on the position of the maximum
we perform $1000$~ fittings of the data points by adding random
Gaussian variables. The standard deviation of the Gaussian noise was
taken to be the error bars in enthalpy.
\bibitem{note2} The $<x^2>$ for DNA was calculated in the same
 fashion as for lysozyme.
\bibitem{note3} The values of $C_P$ are different in each case
with $C_P^{\rm DNA}(T_W)>C_P^{\rm protein}(T_W)$, probably
because DNA hydration level (368\%) is larger than lysozyme hydration
level (156\%) and the fact that the specific heat of water is larger
than protein and DNA.
\bibitem{note4} The difference in the height of the first peak of $g(r)$
for lysozyme and DNA hydration water is due to the different hydration
levels of the two systems.
\end{thebibliography}
\end{document}